\documentclass[Afour,sagev,times, doublespace]{sagej}

\usepackage{moreverb,url}

\usepackage[colorlinks,bookmarksopen,bookmarksnumbered,citecolor=red,urlcolor=red]{hyperref}

\usepackage{graphicx}

\newcommand\BibTeX{{\rmfamily B\kern-.05em \textsc{i\kern-.025em b}\kern-.08em
T\kern-.1667em\lower.7ex\hbox{E}\kern-.125emX}}

\begin{document}

\runninghead{Ranger et al.}

\title{3D ultrasound shear wave elastography for musculoskeletal tissue assessment under compressive load: a feasibility study}

\author{Bryan J. Ranger\affilnum{1}, Kevin M. Moerman\affilnum{2}, Micha Feigin\affilnum{3,4}, Hugh M. Herr\affilnum{5}, Brian W. Anthony\affilnum{3,4}}

\affiliation{\affilnum{1}Department of Engineering, Boston College, Chestnut Hill, MA, USA\\
\affilnum{2}School of Engineering, University of Galway, Galway, Ireland\\
\affilnum{3}Department of Mechanical Engineering, Massachusetts Institute of Technology, Cambridge, MA, USA\\
\affilnum{4}Institute for Medical Engineering and Science, Massachusetts Institute of Technology, Cambridge, MA, USA\\
\affilnum{5}MIT Media Lab, Massachusetts Institute of Technology, Cambridge, MA, USA\\}

\corrauth{Bryan J. Ranger, 245 Beacon Street, Room 516, Department of Engineering, Boston College, Chestnut Hill, MA, USA, 02467}

\email{bryan.ranger@bc.edu}

\begin{abstract}
Given its real-time capability to quantify mechanical tissue properties, ultrasound shear wave elastography holds significant promise in clinical musculoskeletal imaging. However, existing shear wave elastography methods fall short in enabling full-limb analysis of 3D anatomical structures under diverse loading conditions, and may introduce measurement bias due to sonographer-applied force on the transducer. These limitations pose numerous challenges, particularly for 3D computational biomechanical tissue modeling in areas like prosthetic socket design. In this feasibility study, a clinical linear ultrasound transducer system with integrated shear wave elastography capabilities was utilized to scan both a calibrated phantom and human limbs in a water tank imaging setup. By conducting 2D and 3D scans under varying compressive loads, this study demonstrates the feasibility of volumetric ultrasound shear wave elastography of human limbs. Our preliminary results showcase a potential method for evaluating 3D spatially varying tissue properties, offering future extensions to computational biomechanical modeling of tissue for various clinical scenarios.
\end{abstract}


\maketitle

\section{Introduction}
Assessing the mechanical properties of musculoskeletal tissues in a patient-specific manner is imperative for diagnosing various clinical conditions, ranging from muscle injuries \cite{Witvrouw2003} and neuromuscular diseases \cite{Marusiak2010} to athletic training \cite{Kubo2007}. Furthermore, such assessments play a pivotal role in devising effective interventions within rehabilitation medicine, including stretching, strengthening, and spasticity treatments, all of which depend on modifying muscle biomechanical properties \cite{Brandenburg2016}. Nevertheless, quantifying these properties poses challenges due to the musculoskeletal tissue's anisotropic behavior and complex composition of active and passive components \cite{Gennisson2010, bied2021acoustoelasticity,ngo2024unravelling}.

In a manner similar to how clinicians use palpation techniques, medical imaging-based elastography methods have been developed for the assessment and mapping of soft tissue mechanical properties \textit{in vivo} \cite{Sigrist2017}. This approach involves applying a mechanical stimulus to the tissue, followed by imaging the resulting deformation to obtain either qualitative or quantitative stiffness evaluations. One such technique utilized for examining skeletal muscle is magnetic resonance elastography (MRE) \cite{Basford2002, Dresner2001, Heers2003, Jenkyn2003, Uffmann2004}. With MRE, shear waves are generated using an electro-mechanical transducer. Integration with a motion sensitizing gradient in the MR pulse sequence enables the creation of a quantitative stiffness map. Despite its effectiveness, this method is constrained by factors such as cost, accessibility, and various technical limitations such as posture restrictions.

A compelling alternative to MRE is ultrasound shear wave elastography (SWE), a technique now integrated into many clinical ultrasound systems. This approach utilizes acoustic radiation force (ARF) impulses directly from the ultrasound transducer to induce shear waves in the tissue \cite{Arda2011a}. By combining conventional B-mode imaging with speckle tracking methods, real-time tissue displacements caused by shear waves are monitored, creating a map of shear wave velocities. These velocities directly correlate with tissue shear modulus according to the equation:  $\mu = \rho c^2$, where $\mu$ is the shear modulus, $\rho$ is tissue density, and $c$ is shear wave velocity \cite{Nightingale2003}. Unlike other elastography techniques, such as strain imaging which assesses relative stiffness between tissues, SWE offers a quantitative modulus value on an absolute scale \cite{Brandenburg2014UltrasoundStiffness}. 

Recently, SWE has been applied to numerous musculoskeletal applications \cite{Taljanovic2017}. Several studies have illustrated ultrasound SWE's capability to gauge muscle stiffness across various clinical scenarios \cite{Bernabei2020ShearForce, Liu2020InElastography, Liu2021QuantitativeAges, Olchowy2020PotentialStiffness, Sarabon2019UsingMuscle}. The technique's efficacy has been assessed for measuring human muscle stiffness with different transducer angles and orientations \cite{Miyamoto2015ValidityMuscle}. Numerous advances have demonstrated more nuanced characterization of the elastic, in-compressible, and transversely isotropic properties of muscle using 3D SWE imaging \cite{knight2020demonstration, knight2021full, trutna2020viscoelastic}. Studies have also explored the quantification of muscle fiber orientation and rotational measurements for SWE \cite{jin2021quantification,paley2023repeatability, paley2024rotational}. Despite these advancements, these studies did not provide a mechanism to characterize the effects of an externally applied load to the imaged body segment on mechanical tissue properties. 

While SWE holds promise for quantifying musculoskeletal tissue stiffness, there are several limitations because of its inherent operator dependence. Specifically, due to the non-linear behavior of biological soft tissue, when a transducer deforms the tissue there is a resulting change in stiffness \cite{VanLoocke2009} and thus a change in SWE results – this has been shown in various studies including those focused on breast \cite{Weismann2011}, liver \cite{Porra2015}, kidney \cite{Syversveen2012a}, and muscle \cite{Kot2012}. In each of these studies, the operator’s transducer force directly affected the investigated tissue properties, leading to variances in the resulting stiffness modulus. For muscle specifically, Kot el al.\cite{Kot2012} recommended that when SWE is collected, that an examiner should use the lightest transducer pressure possible in order to improve accuracy. Similarly, Cortez et al.\cite{Cortez2016} showed high reliability in shear wave velocity measurements under stringent experimental conditions but only moderate reproducibility in scenarios resembling clinical practice. Consequently, there is a demand for reliable methods to evaluate muscle tissue composition\cite{Correa-de-Araujo2017a}, and substantial efforts should be directed towards standardizing measurement procedures to ensure the acquisition of meaningful data for diagnostic assessment\cite{Ryu2017a}.

In efforts to reduce operator dependence, numerous research studies have investigated incorporating a reference layer within a controlled setup to quantify tissue elastic properties and mitigate variations in transducer-applied force \cite{Selladurai2019TowardsAssessment}. Despite these advancements, SWE often confines elastographic measurements to a specified 2D region of interest (ROI), restricting the analysis of volumetric pathological features or tissue structures. The challenge of applying a transducer-independent load hampers ultrasound's capability to assess the 3D mechanical response of tissue to an external force. Accordingly, there is a need for a volumetric ultrasound dataset encompassing tissue geometry and biomechanical properties in multiple dimensions. Such a dataset would afford clinicians a cost-effective means of obtaining a comprehensive spatial understanding of tissue properties for diagnostic purposes and interventions\cite{Yin2018}.

Integrating mechanical tissue property data with volumetric ultrasound imaging, coupled with computational modeling such as finite element analysis (FEA), holds promise for refining ultrasound-based constitutive modeling of biological soft tissue \cite{fougeron2020combining}. A specific application of this concept is computational prosthetic socket design. Prosthetic sockets are the cup-like interfaces that connect a residual limb to a prosthesis, and are traditionally handcrafted through a plaster casting process. While recent advancements leverage computer-aided design and manufacturing (CAD/CAM) for this process\cite{herr2021quantitative, Sengeh2013, Sengeh2016}, it often overlooks internal tissue distributions and mechanical properties, or relies on costly scanning tools. While FEA can assess large strain and non-linear elastic formulations, it lacks the ability to derive spatially varying maps of constitutive behavior. Computational biomechanics models can simulate tissue response to various loading conditions, such as mechanical forces or deformations. By integrating experimental data from ultrasound elastography into these models, researchers can improve their ability to predict tissue behavior under different loading scenarios, which is crucial for applications such as biomechanical simulations for different loading conditions in the prosthetic socket design\cite{ranger2023constitutive}. This underscores the significant potential of 3D ultrasound SWE in advancing prosthetic technology.

In summary, there is an unmet need in ultrasound SWE to capture 3D images and to biomechanically characterize muscles under diverse loading conditions. To address this gap, we present a feasibility study employing ultrasound SWE in a water tank. First, we demonstrate the acquisition of consistent shear wave velocity measurements from both a calibrated phantom and human limbs without direct transducer contact with the imaged object or body segment. In this evaluation, we obtained 2D images of the phantom at varying distances from the transducer to demonstrate accuracy and precision. Additionally, we present 2D SWE results of human limbs under different compressive loads, along with the creation of 3D shear wave velocity maps of the limb while applying an external transducer-independent load, providing a feasbility assessment of the concept \textit{in vivo}.

\section{Materials and Methods}

Measurements in this study were conducted employing a 9 $MHz$ linear array ultrasound transducer system (9L-D Linear Array Probe, LOGIQ E9, GE, Niskayuna, NY) with integrated SWE capabilities. The ROI was configured in a set rectangular shape and its spatial position could be manually adjusted. Both a calibrated phantom and human limbs were imaged, and a 3D scan was executed to illustrate the viability of volumetric elastography under varying compressive loads. All data and image processing were executed using MATLAB (R2017A, Mathworks, Natick, MA), complemented by the open-source Geometry and Image-Based Bioengineering Add-On (GIBBON) toolbox\cite{Moerman2018}.

\subsection{2D Phantom Scans}

A calibrated phantom (Model 039, Computerized Imaging Reference Systems, Inc. (CIRS), Norfolk, VA) of known shear wave velocity (4 $m/s$) and made of homogeneous material with no inclusions, was scanned using ultrasound shear wave elastography. Images were acquired at incremental distances between the transducer surface and phantom. For each distance, at least 20 readings were taken and averaged together to assess accuracy and precision.

With the phantom submerged in the water tank imaging setup (\textbf{Figure 1A}), shear wave velocity images were acquired. Deionized water was used in the tank to minimize the effect of sound speed gradients. To mitigate any potential motion artifacts, the transducer was securely clamped to a ring stand and maintained in a stationary position. Data were systematically collected at 0.5 $cm$ incremental distances (starting at 0 $cm$) to analyze the measurement accuracy and precision as a function of distance between the transducer and the phantom surface. The aim of this experiment was to verify that measurements could be acquired at different locations even if the physical distance between probe and imaged object were to vary. 

\subsection{2D Human Limb Scans}

Two adult subjects were recruited following a procedure approved by the MIT Committee on the Use of Humans as Experimental Subjects (COUHES), and informed written consent was obtained. Both subjects were healthy male adults – age: mean 27.5 (SD 2.1), BMI: mean 22.7 (SD 2.4). 2D measurements were acquired for each subject using both the gel approach and water tank approach.

\textbf{Figure 1} shows the experimental setup for the 2D human limb scans – \textbf{Figure 1B} depicts the setup using gel, while \textbf{Figure 1C} shows the water tank setup. As depicted, the transducer is positioned at the subject’s calf, and their limb was in the same position for each reading (i.e., foot flat and resting on the ground with limb perpendicular to the ground). The subject was seated to minimize motion and ensure their comfort. When utilizing gel for measurements, a minimum of 0.5 $cm$ of gel was applied between the transducer and the limb to mitigate the impact of the transducer force on the tissue. 
Shear wave velocity measurements (in $m/s$) were obtained in both an unloaded state and across three distinct compressive states, all without the transducer making contact with the limb. The rectangular region of interest (ROI) was manually centered in the same tissue layer location of the limb for each scan. The ROI selection aimed to minimize the depth of measurement based on findings in other studies, which demonstrated consistent readings at less than 4 $cm$ depth \cite{alfuraih2018effect}, and to mitigate any effects associated with the presence of bone boundaries \cite{ewertsen2016evaluation}. Additionally, the selection of the ROI for shear wave elastography (SWE) maximized the horizontal distance, a factor shown in some studies to yield more reproducible results \cite{ruby2019confounders}. 

Compressive states were induced using compression garments with different grades (Allegro Compression Hosiery, BrightLife Direct, Washington, DC) at 8-15 $mmHg$, 15-20 $mmHg$, and 20-30 $mmHg$. These medical-grade socks were composed of 88\% nylon and 12\% spandex, demonstrating no discernible impact on image or signal quality when worn. Given the transversely isotropic nature of muscle tissue, shear wave velocity measurements were captured in both longitudinal and transverse orientations for both gel and tank scans. For each compressive state, two sets of measurements, each comprising at least 20 readings, were acquired.

\subsection{3D Human Limb Scans}

3D SWE imagery of a subject’s limb was collected in in a water tank setup, shown in \textbf{Figure 1D}. This system has been described in detail in previous publications from our group \cite{Ranger2015,Ranger2016a,Ranger2017b,Ranger2019}. In summary, the system consists of a water tank with a ring bearing mounted on top. A custom 3D-printed mount secures the ultrasound transducer to the rotating portion of the ring bearing, and allows for circumferential rotation of the transducer around the limb at a fixed radius. To conduct these scans, shear wave elastography (SWE) images were gathered at 10-degree intervals around the limb, with the transducer oriented in the longitudinal direction. At each degree increment, at least 5 measurements were taken and averaged together. Compression states were induced using sheer support compression garments at 8-15 $mmHg$, 15-20 $mmHg$, and 20-30 $mmHg$.

\subsection{Data Analysis}

Shear wave velocity maps were extracted from the DICOM data and imported into MATLAB for analysis. To examine spatially varying properties, mean and standard deviation values were computed for all measurements taken in each orientation. For the volumetric visualization of 3D-collected data, images were initially arranged radially in space according to the orientation in which they were acquired. Subsequently, these images were transformed into scattered data and sampled onto a regular grid. Natural-neighbor interpolation \cite{Sibson1980}, as implemented using the \textit{scatteredInterpolant} class in MATLAB, was used to process both the B-mode data as well as the SWE data. For visualization, water was segmented out of the 3D rendering using the \textit{imx} function in the GIBBON MATLAB toolbox.

\section{Results}

\textbf{Table 1} presents mean and standard deviation (SD) values for shear wave velocity maps obtained at incremental distances from the phantom in the water tank. The mean shear wave velocity consistently aligns within $\pm$ 0.2 $m/s$ of the expected calibrated phantom value (4 $m/s$) up to distances of 6 $cm$. This affirms the capability of acquiring consistent shear wave velocity measurements in a water tank setup, showcasing a broad range of transducer distances compared to a conventional gel approach where the probe is making direct contact with the imaged object or body segment. The standard deviation remains within the same range until 8 $cm$, underscoring the precision of shear wave velocity measurements across a wide span of transducer distances.

\textbf{Figure 2} depicts example ultrasound images, in both longitudinal and transverse orientations, acquired of a subject’s leg under the four compressive states. The color map of shear wave velocity measurements rises with increasing compression, as expected. \textbf{Table 2} and \textbf{Table 3} show mean and standard deviation values for the shear wave velocity maps for the two research subjects. Each plot includes measurements from both the gel and water tank approaches for comparison. With the exception of standard deviation measurements for the gel approach in Subject 2, both mean shear wave velocity and standard deviations exhibit approximately a twofold increase from the unloaded state to the highest loaded (20-30 $mmHg$) state. As shown in these tables, measurable changes in mean shear wave velocity may be detected when applying a transducer-independent external compressive load to the limb, and measurements taken from the standard gel approach are comparable to those acquired in the water tank. Thus, \textit{in vivo} shear wave velocity measurements of human muscle as performed in a water tank demonstrates the feasibility for scanning around a body segment to create a 3D image.  

Extending on the results of the calibrated phantom, 2D images of human limbs (\textbf{Figure 2}) under four different compressive loads are presented. For each limb, and in both the transverse and longitudinal orientations, shear wave velocity measurements increased as a function of applied external load. Measurements (\textbf{Table 2} and \textbf{Table 3}), ranged from 1.69$\pm$0.58 to 1.89$\pm$0.71 $m/s$ in the longitudinal orientation, and from 1.77$\pm$0.59 to 1.82$\pm$0.58 $m/s$ in the transverse orientation for mean shear wave velocity for each subject. These values align within a comparable range to those reported in other studies that have gathered data on human limbs – for example, Cortez \textit{et al} reported shear wave velocity measurements of the gastrocnemius medialis ranged from 1.89$\pm$0.32 to 2.38$\pm$0.58 $m/s$ in the longitudinal orientation, and from 1.54$\pm$0.22 to 1.94$\pm$0.29 $m/s$ in the transverse orientation \cite{Cortez2016}. 

Images collected at angular increments around the limb were stitched together in Matlab using Natural-neighbor interpolation. \textbf{Figure 3} illustrates the generation of the volumetric image dataset. In \textbf{Figure 3A} B-mode images obtained at 10-degree intervals around a subject’s limb positioned in 3D space are presented. In \textbf{Figure 3B}, corresponding SWE images captured at the same 10-degree intervals around the subject’s limb in 3D space are depicted. The collected images are then re-gridded and undergo natural-neighbor interpolation in 3D. 
The SWE results are fused with the B-mode images, resulting in a 3D visualization of SWE and general anatomy shown in \textbf{Figure 4}. Similar to the 2D results, a measurable increase in shear wave velocity is noted as load increases, as shown in \textbf{Figure 5A}. In this plot, spatially varying properties are portrayed at different circumferential locations around the limb and at distinct anatomical landmarks. \textbf{Figure 5B} further explores this notion by normalizing each compressive state to the original unloaded state – in this plot, shear wave velocity values are elevated in tissue regions near the tibia and fibula bones. 

\section{Discussion}

This study demonstrates feasibility for conducting  3D SWE in a water tank, which enables the imaging of a limb in an undeformed state as well as under different external compression conditions. By demonstrating the feasibility of obtaining a 3D ultrasound SWE dataset under varying loading conditions, that contains both tissue morphology and biomechanics in multiple dimensions, we highlight its potential clinical relevance in musculoskeletal ultrasound. This approach offers a cost-effective means to attain a comprehensive spatial understanding of limb tissue structures and spatially varying tissue properties, which may ultimately assist with improved diagnosis, computational biomechanical tissue modeling in areas such as prosthetic socket design, and clinical intervention planning.

First, to demonstrate the accuracy of using shear wave velocity as a measurement tool in a water tank, 2D images were acquired of a calibrated phantom at varying distances from the transducer in a water tank (non-contact) setup. Accurate measurements of the calibrated phantom were achieved. An important observation from these results is that the water tank setup allowed for measurements to be taken at further distances away from the phantom, as compared to the conventional gel approach where the probe is making contact. We also note that shear wave velocity measurements increased as distance between the transducer and phantom increased in the water tank. This is likely attributable to how the sound waves emitted by the transducer, in particular the ARF impulse `push' beams,  diffract at these larger distances. More specifically, diffraction of the ARF impulse `push' and `tracking' beams for SWE spread out as they propagate through the water, with this effect becoming more pronounced at longer distances. As the distance between transducer and phantom increased over the 9 $cm$ range in this study, it is likely that the tracking beams intersected with the region affected by the ARF impulse `push' and interfered with the accurate measurement of shear wave propagation speed, which can ultimately lead to biased estimates of stiffness. In summary, this preliminary study successfully obtained consistent measurements of shear wave velocity of a calibrated phantom in a water tank, showcasing the capability to collect SWE data without direct transducer contact with the imaged body and across a broad range of transducer distances. This verifies the approach in 3D scanning scenarios, particularly where the ultrasound probe follows a fixed scanning path and the distance between the imaged body segment and transducer may vary.

We then conducted 2D SWE ultrasound experiments on human lower limbs in a water tank setup. Consistent shear wave velocity measurements were acquired while the transducer was not contacting the limb by performing the scan in a water tank, and 3D shear wave velocity measurements were acquired while an external load (i.e., with compression garments, and not from the transducer) is applied to the imaged segment of a limb. Because the water acts as the acoustic coupling medium, no gel is needed, and the transducer does not need to make direct contact with the limb. Though water acts as the coupling medium, we note that the speed of sound of deionized water differs from the estimated speed of sound for coupling gel which is adjusted for use on human tissue. As such, it is possible that some biases exist when quantifying shear wave velocities in the water tank. Additional studies on a larger cohort of patients may explore this further to determine whether there is significance related to these potential biases. When comparing SWE values of the limb collected in our study to the literature, we note that values are within comparable range. We acknowledge that this is not a fair direct comparison, as we expect variability to be  higher in our study with limited experimental subjects compared to the results of studies that have a larger sample size. 

To extend on the 2D data collection, we utilized water tank scanning system to collect 3D SWE of human lower limbs \cite{Ranger2019}. Utilizing Natural-neighbor interpoloation methods, and re-gridding, we fused the SWE and B-mode data to achieve a 3D visualization of SWE under different loading conditions. By scanning circumferentially around the limb, we are able to observe spatially varying properties of the tissue. Anisotropy, influenced by the directional alignment of muscle fibers, leads to differential stiffness along different orientations relative to these fibers. Compression during the scan affects tissue deformation, potentially resulting in increased stiffness in regions subjected to higher compressive forces, such as areas near bones or under greater load, while regions experiencing less compression may display lower stiffness values. At locations near the bones, we observed increased strain produced in the tissue because of increasing load, thus, our measurements agree with our expectations. To the author’s knowledge, this is the first demonstration of volumetric SWE measurements collected in a water tank system while varying a transducer-independent compressive load on a limb. Though this represents just one section of the limb, these preliminary results demonstrate feasibility to perform a full volumetric SWE scan of a limb in a non-contact water tank system under various loading conditions, and to identify spatially varying properties. Such 3D SWE data of the limb under different loading conditions is particularly relevant for biomechanical modeling of the tissue, in areas like computational prosthetic socket design.

One of the primary limitations of this study is the frame rate of the GE Logiq E9 system; to acquire one shear wave velocity map, the frame rate was approximately 1 frame per second. Given this, scan time was limited since multiple measurements at several orientations were needed to evaluate consistency, and scanning was only performed on one section of the limb. By utilizing a system with a higher frame rate, this issue may be resolved in future related studies to allow for full 3D limb imaging and should not pose significant challenges under more realistic clinical conditions. An additional limitation of this study was that the compression garments represented a range of possible pressures since it could vary based limb size and different anatomical locations. This option was justified for the preliminary nature of our work, but future studies could explore other means of providing a more uniform and specific pressure field to the imaged body segment to alleviate this limitation. Another constraint in this study pertained to the selection of the ROI, which focused on superficial tissue layers for SWE measurements. This choice was justified given that one of the primary areas of application is in developing models for computational biomechanics of tissues for prosthetic interface design, with a specific interest in understanding the impact of mechanical forces onto the tissue. Given that the most discernible changes related to varying interface conditions in our experimental setup would manifest in the superficial layers of the tissue, this study focused on measurements in these tissue regions. The limited scan rate further restricted our examination to superficial tissue layers. This, in combination with the slow scan rate, limited the data collection to superficial tissue layers. Future work beyond this feasibility study should take into account measurements of additional muscle layers.

Limb scans were performed while the subject rested their limb flat on the bottom of the imaging tank. Because of this, some muscle groups may not have been in a completely passive state.  This positioning is justified since it minimized motion during the scan and allowed for consistent spatial positioning between scans. To ensure investigation of passive properties, future iterations of this work may include re-envisioning how the subject is positioned, as well as implementing appropriate motion compensation algorithms \cite{Ranger2015,Ranger2016a,Ranger2017b,Ranger2019}, to alleviate this potential source of error. In addition, it is also important to note that the scanning device was constructed as a research tool that will allow for analysis of the elastic properties of the limb under load, and would require further iterative design to ultimately translate to the clinic.
3D SWE data has significant potential to be incorporated into the field of computational modeling of biological soft tissue. Recently, or lab published on the use of ultrasound SWE for constitutive modeling of residual limb tissue utilizing shear wave measurements taken at different force increments during an indentation experiment \cite{ranger2023constitutive}. In this work, we demonstrated that SWE allowed for modeling of spatially varying tissue properties, in a low cost and accurate way. The work presented in this paper offers a direct extension of these ideas but to 3D. Therefore, by providing patient-specific measurements of tissue stiffness in 3D at different compression states, this data could inform more realistic constitutive modeling and analysis of tissue non-linearities \cite{Avril2010}. For example, this data may be used for 3D constitutive modeling of limb biomechanics by providing patient-specific information on spatially varying tissue properties, and could have broader applications in impact biomechanics \cite{Forbes2005}, rehabilitation engineering \cite{Linder-Ganz2007}, surgical simulation \cite{Famaey2008}, and prosthetic socket design \cite{Sengeh2016,Moerman2016}. 

The results of this proof-of-concept study also have broad implications for clinical imaging. Given that this approach enables 3D elastography of a limb and mitigates several elements of operator dependency, it facilitates further quantitative research across various musculoskeletal applications. If combined with additional viscoelastic tissue property analysis, future work could include longitudinal studies of disease progression, more accurate analysis of muscle state and contraction ability, large-scale 3D elastography, and detailed comparative studies between patients. Further, these techniques could also be applied to other ultrasound tomography based systems. For example, in 3D tomographic imaging for breast cancer \cite{Duric2010In-vivoInstitute}, elastography could potentially be explored as a method for tumor assessment.  

In summary, the preliminary results of this feasibility study demonstrate encouraging advances in non-contact SWE and warrant further investigation. A future study from our groups will explore the direct incorporation of SWE measurements into the computational analysis of soft tissue biomechanics, with a particular focus on residual limbs and applications to prosthetic socket design. 

\section{Conclusion}

The feasibility results presented in this paper represent a significant advancement for SWE as it pertains to musculoskeletal health, particularly in the areas of computational tissue biomechanics and quantitative prosthetic socket design. We demonstrate that consistent SWE measurements of a calibrated phantom and human limb may be acquired and that 3D \textit{in vivo} elastography data of the limb may be collected under different loading conditions in a water tank setup. This lays important groundwork for future quantitative MSK research and clinical studies in which limbs may be imaged in 3D using SWE without the transducer making direct contact with the limb. 




\begin{acks}
The authors would like to thank the students and staff of the MIT Device Realization Lab and MIT Media Lab Biomechatronics Group for their support related to this project; in particular, Dana Solav, Rebecca Zubajlo, Xiang Zhang, and Jonathan Fincke. We also thank Stephanie Ku who assisted with experimental setup schematics. 
\end{acks}

\section{Funding}
This work was supported in part by the National Science Foundation Graduate Research Fellowship Program (NSF-GRFP) and the MIT Media Lab Consortia.

\bibliographystyle{SageV} 
\bibliography{references}

\newpage
\onecolumn

\section{Figures}

\begin{figure*}[h]
\captionsetup{justification=justified}
\includegraphics[width=\textwidth]{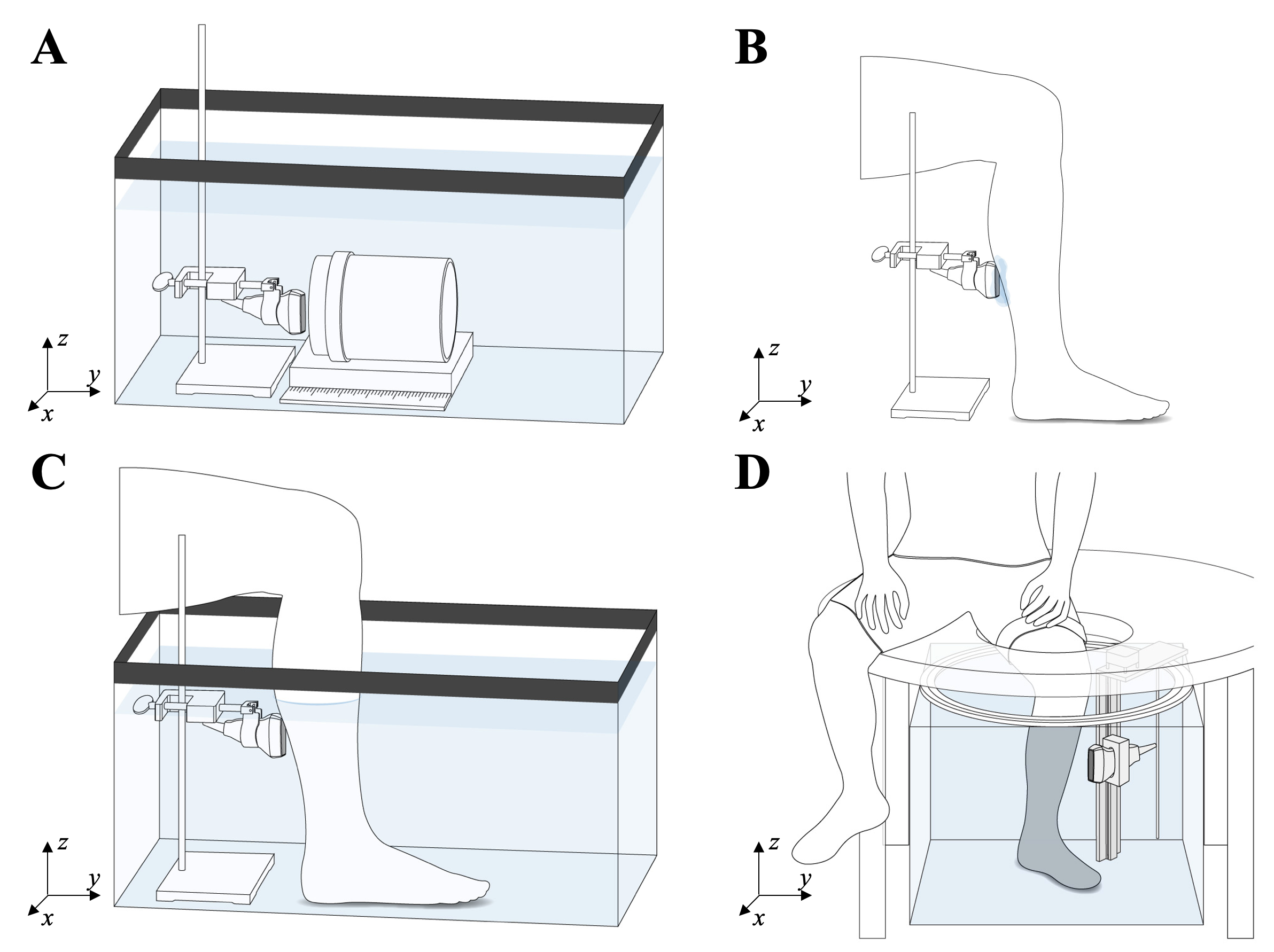}
\caption{Experimental Setups. (A) Setup for scanning the calibrated phantom in the water tank. A ring stand was used to secure the ultrasound transducer into a fixed position. The phantom may be moved at incremental distances away from the ultrasound transducer in the tank. (B) Setup for scanning a limb with the standard gel approach. The ultrasound transducer is fixed to a ring stand facing toward the limb. A layer of ultrasonic coupling gel is placed between the transducer and limb surface.  (C) Setup for scanning the limb in the water tank. A ring stand is used to secure the ultrasound transducer. Distance between transducer and limb is held constant between each scan. (D) Using a setup previously described \cite{Ranger2019}, a research subject’s limb was scanned to collect 3D shear wave elastography data. The setup consists of a water tank with a ring bearing mounted on top. A custom 3D-printed mount secures the ultrasound transducer to the rotating portion of the ring bearing, thus allowing for circumferential rotation of the transducer around the limb at a fixed radius. Shear wave elastography imagery was collected at 10-degree increments around the limb in the longitudinal directions. The subject is situated above the tank and is asked to submerge their limb into the tank and place their foot flat on the bottom to minimize motion.}
\label{fig:Figure_1}
\end{figure*}

\begin{figure*}[h]
\captionsetup{justification=justified}
\includegraphics[width=\textwidth]{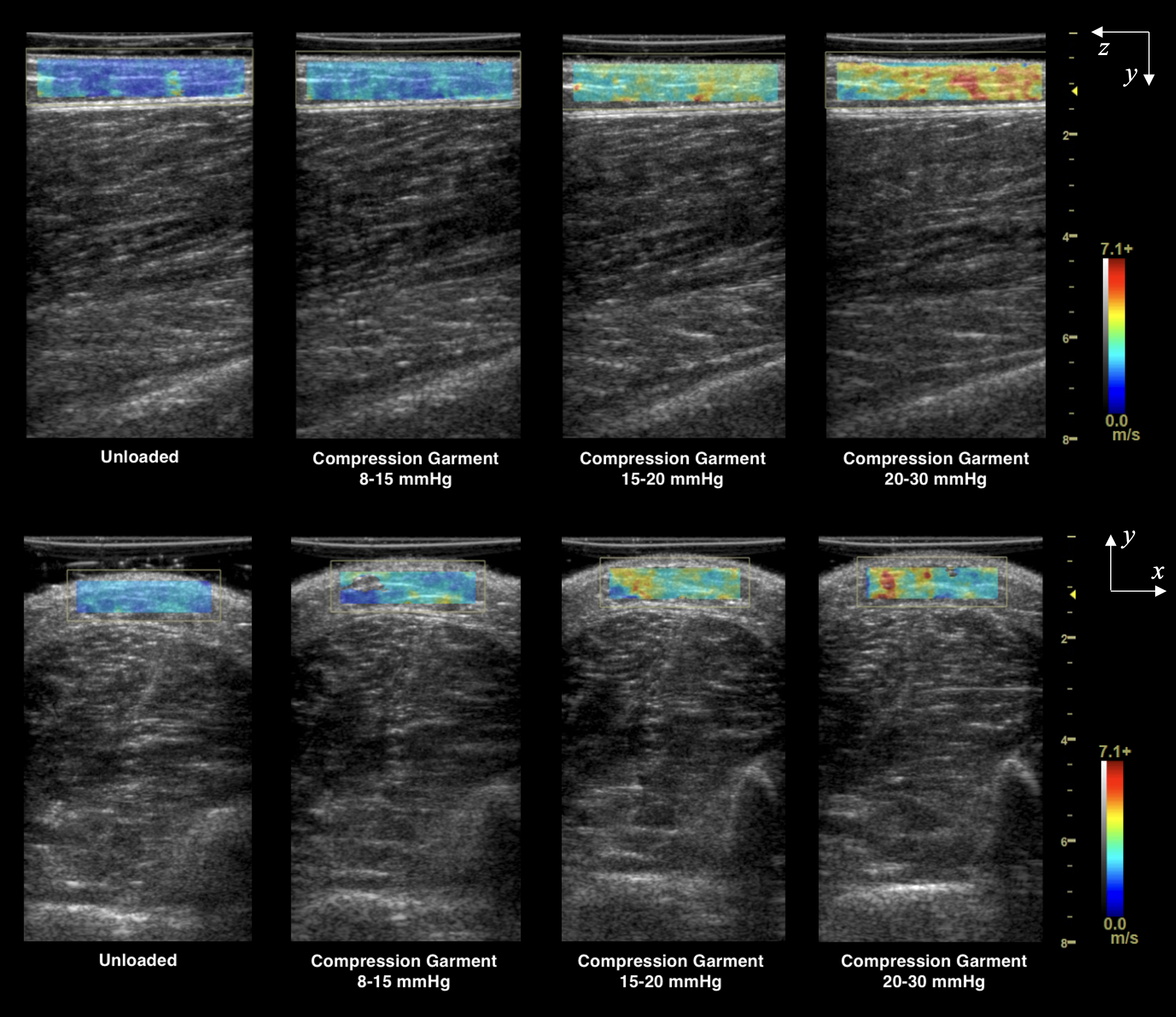}
\caption{Ultrasound images collected in the longitudinal direction (top row) and transverse direction (bottom row) Subject 1's leg under four compressive states: uncompressed, 8-15 mmHg grade compression sock, 15-20 mmHg grade compression sock, and 20-30 mmHg grade compression sock. As shown, the heat map of shear wave velocity measurements increases with increasing compression. The ultrasound images of the lower limb contain superficial tissues, gastrocnemius, and soleus muscles.}
\label{fig:Figure_2}
\end{figure*}

\begin{figure*}[h]
\captionsetup{justification=justified}
\includegraphics[width=\textwidth]{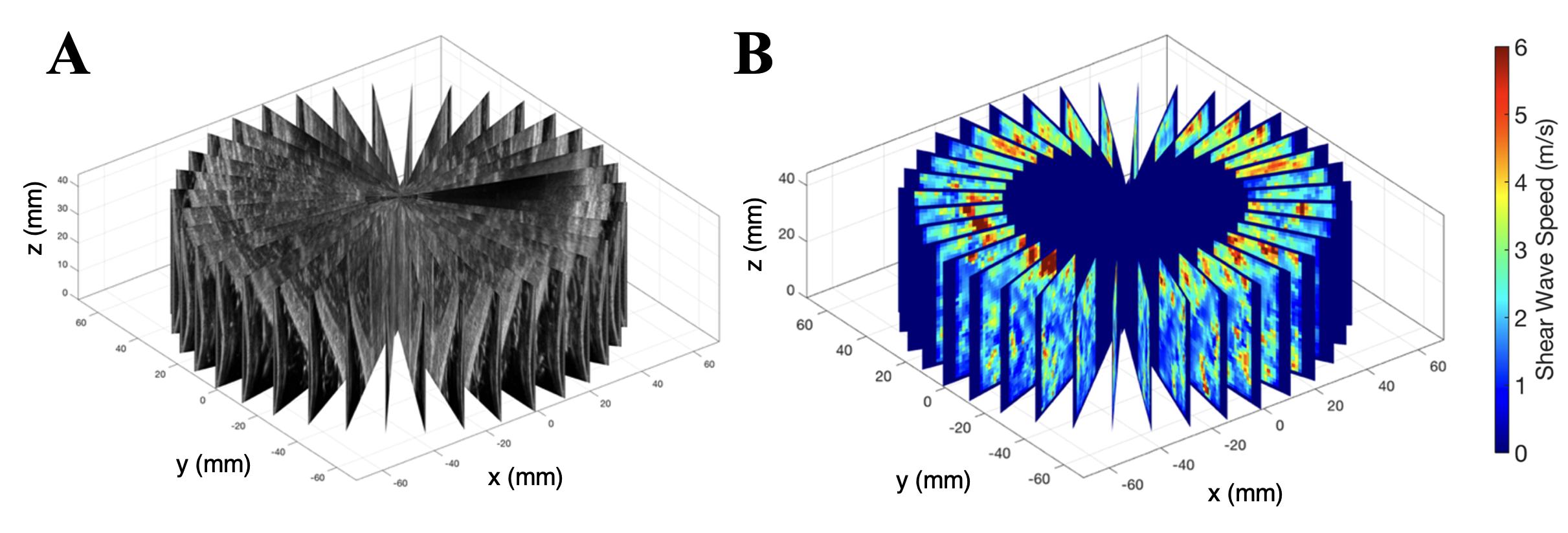}
\caption{Creation of a volumetric image data set. (A) B-mode images collected at 10-degree increments around a subject’s limb placed in 3D space. (B) SWE images collected at 10-degree increments around a subject’s limb placed in 3D space.}
\label{fig:Figure_3}
\end{figure*}

\begin{figure*}[h]
\captionsetup{justification=justified}
\includegraphics[width=\textwidth]{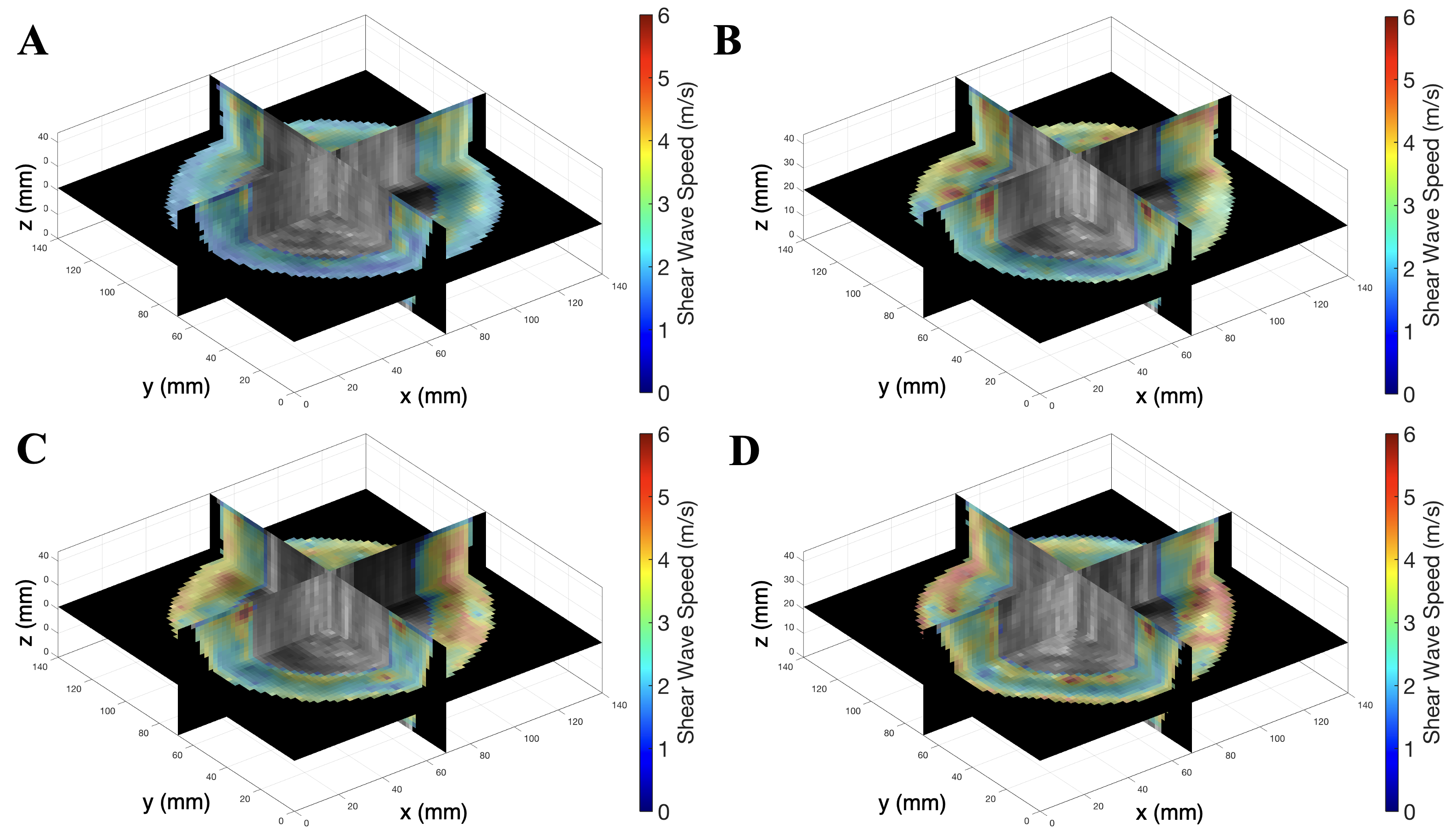}
\caption{Volume results showing 3D changes in shear wave velocity, overlaid on B-mode data, at varying compressive loads. (A) Unloaded. (B) 8-15 mmHg compression garment. (C) 15-20 mmHg compression garment. (D) 20-30 mmHg compression garment.}
\label{fig:Figure_4}
\end{figure*}

\begin{figure*}[h]
\captionsetup{justification=justified}
\includegraphics[width=\textwidth]{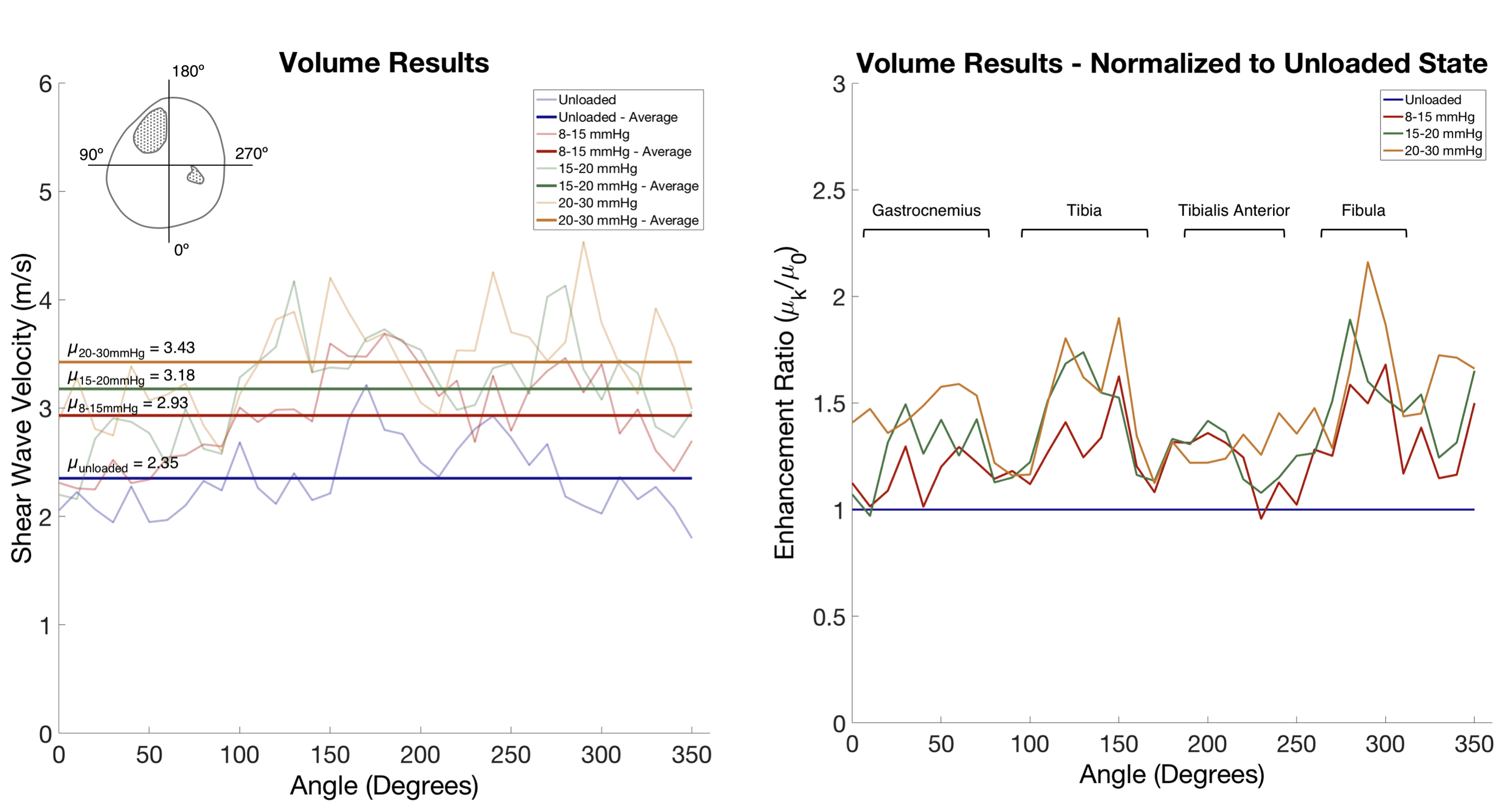}
\caption{(A) Plot of average shear wave velocity values collected at the 10-degree increments around the limb. The raw mean value collected at each of the four compression states is plotted at each angle, and the total average for each compression state is plotted as a horizontal line with mean value noted. (B) Plot of average shear wave velocity values collected at the 10-degree increments around the limb, normalizing the data for each compression state to the unloaded state. Elevated mean shear wave velocity values in bony regions of the limb are noted as compared to soft tissue regions, demonstrating spatially varying tissue properties.}
\label{fig:Figure_5}
\end{figure*}

\newpage
\onecolumn

\section{Tables}

\begin{table}[h]
\small\sf\centering
\caption{Mean and standard deviation shear wave velocity values of the calibrated phantom (4 m/s) in water as a function of distance from transducer}
\begin{tabular}{ll}
\toprule
Distance (cm) & Shear Wave Velocity (m/s) of Phantom\\
\midrule
0.0 &  3.81$\pm$0.09 \\ 
0.5 & 3.93$\pm$0.20  \\ 
1.0 & 3.91$\pm$0.18 \\ 
1.5 & 3.93$\pm$0.16 \\ 
2.0 & 3.94$\pm$0.18 \\ 
2.5 & 3.88$\pm$0.16 \\ 
3.0 & 3.87$\pm$0.17 \\ 
3.5 & 3.94$\pm$0.17 \\ 
4.0 & 3.94$\pm$0.20 \\ 
4.5 & 3.92$\pm$0.19 \\ 
5.0 & 4.07$\pm$0.20 \\ 
6.0 & 4.15$\pm$0.20 \\ 
6.5 & 4.31$\pm$0.20 \\ 
7.0 & 4.29$\pm$0.25 \\ 
8.0 & 4.34$\pm$0.35 \\ 
9.0 & 4.45$\pm$0.64 \\ 
\bottomrule
\end{tabular}\\[10pt]
\label{table1}
\end{table}

\begin{table}[!ht]
\centering
\caption{Mean and standard deviation shear wave velocity values for the two human subjects in gel and water and in longitudinal orientations}
\begin{tabular}{|c|c|c|c|c|}
\hline
{\textbf{Compression}} & \multicolumn{2}{l|}{\textbf{Subject 1 - Longitudinal (m/s)}} & \multicolumn{2}{l|}{\textbf{Subject 2 - Longitudinal (m/s)}} \\ \cline{2-5} 
& \textbf{Gel} & \textbf{Water} & \textbf{Gel} & \textbf{Water} \\ \hline
Unloaded & 1.69$\pm$0.58 & 1.58$\pm$0.61 & 1.89$\pm$0.71 & 1.67$\pm$0.47 \\ \hline
8-15 mmHg & 2.00$\pm$0.73 & 2.01$\pm$0.63 & 2.27$\pm$0.70 & 2.18$\pm$0.83 \\ \hline
15-20 mmHg & 2.73$\pm$0.04 & 2.94$\pm$0.79 & 2.97$\pm$0.79 & 2.64$\pm$0.91 \\ \hline
20-30 mmHg & 3.36$\pm$1.37 & 3.66$\pm$1.19 & 3.23$\pm$0.94 & 2.90$\pm$0.81 \\ \hline
\end{tabular}
\label{table2}
\end{table}

\begin{table}[!ht]
\centering
\caption{Mean and standard deviation shear wave velocity values for the two human subjects in gel and water and in transverse orientations}
\begin{tabular}{|c|c|c|c|c|}
\hline
{\textbf{Compression}} & \multicolumn{2}{l|}{\textbf{Subject 1 - Transverse (m/s)}} & \multicolumn{2}{l|}{\textbf{Subject 2 - Transverse (m/s)}} \\ \cline{2-5} 
& \textbf{Gel} & \textbf{Water} & \textbf{Gel} & \textbf{Water} \\ \hline
Unloaded & 1.82$\pm$0.58 & 1.91$\pm$0.57 & 1.77$\pm$0.59 & 1.67$\pm$0.50 \\ \hline
8-15 mmHg & 2.45$\pm$0.96 & 2.31$\pm$0.91 & 2.34$\pm$1.01 & 1.90$\pm$0.56 \\ \hline
15-20 mmHg & 3.10$\pm$1.22 & 2.97$\pm$1.02 & 2.95$\pm$1.15 & 2.77$\pm$0.86 \\ \hline
20-30 mmHg & 3.56$\pm$1.43 & 3.62$\pm$1.37 & 3.45$\pm$1.28 & 3.26$\pm$1.11 \\ \hline
\end{tabular}
\label{table3}
\end{table}

\end{document}